\documentclass[reprint,
 amsmath,amssymb,
 aps,
]{revtex4-2}

\usepackage{graphicx}
\usepackage{dcolumn}
\usepackage{bm}
\usepackage{xcolor}
\usepackage{mathptmx}
\usepackage{graphicx}
\usepackage{dcolumn}
 
\usepackage{bm}
\usepackage{amssymb}
\usepackage{comment}
\usepackage{hyperref}

\begin{document}

\title{Abrupt X-to-O-wave structural field transition in presence of anomalous dispersion}

\author{Layton A. Hall$^{1}$}
\author{Ayman F. Abouraddy$^{1,*}$}
\affiliation{$^{1}$CREOL, The College of Optics \& Photonics, University of Central~Florida, Orlando, FL 32816, USA}
\affiliation{$^*$Corresponding author: raddy@creol.ucf.edu}

\begin{abstract}
All linear, propagation-invariant, paraxial pulsed beams are spatiotemporally X-shaped (conical waves) in absence of group-velocity dispersion (GVD), or in presence of normal GVD. It is known, however, that such conical waves become O-shaped in presence of anomalous GVD, resulting in a field profile that is circularly symmetric in space and time. To date, experiments generating conical waves in which the wavelength of a high-energy pump laser is tuned across the zero-dispersion wavelength of a nonlinear medium have not revealed the expected X-to-O-wave structural field transition. We report here unambiguous observation of a \textit{fixed-wavelength} X-to-O-wave structural field transition occurring in linear dispersion-free wave packets in the anomalous GVD regime -- without needing to change the sign or magnitude of the GVD. Instead, by tuning the group velocity of a space-time wave packet (STWP) across a threshold value that we call the `escape velocity', we observe an \textit{abrupt} transition in the STWP from an O-shaped to an X-shaped spatiotemporal profile. This transition is associated with an abrupt change in the associated spatiotemporal spectrum of the STWP: from closed elliptical spatiotemporal spectra below the escape velocity to open hyperbolic spectra above it. These results may furnish new opportunities for engineering the phase-matching conditions in nonlinear and quantum optics. 
\end{abstract}


\maketitle

\section{Introduction}

The interplay between diffraction, group-velocity dispersion (GVD), and optical nonlinearity can permit pulsed beams (or wave packets) to propagate invariantly, free of diffraction and dispersion, in the form of spatiotemporal solitons \cite{DTrapanii98PRL,Liu00PRL,Liu00PRE,Wise02OPN}. However, propagation-invariant wave packets can exist even in \textit{linear} dispersive media when endowed with prescribed space-time coupling \cite{Porras03PRE2,Porras03OL,Longhi04OL,Porras04PRE,Christodoulides04OL,Mills12PRA}. In other words, only certain non-separable spatiotemporal profiles are compatible with linear propagation invariance in a dispersive medium \cite{FigueroaBook14}. In free space, all propagation-invariant wave packets have an X-shaped spatiotemporal profile (conical waves) \cite{Saari97PRL,Grunwald03PRA,Jedrkiewicz06PRL,Jedrkiewicz07PRA,Bowlan09OL,Turunen10PO}. In presence of \textit{normal} GVD, propagation-invariant wave packets are either X-shaped \cite{Sonajalg97OL,Malaguti09PRA,Jedrkiewicz13OE}, which are associated with open \textit{hyperbolic} spatiotemporal spectra having -- in principle -- no upper bound on their bandwidth, or are separable with respect to space and time \cite{Liu98JMO,Hu02JOSAA,Lu03JOSAA}. In presence of \textit{anomalous} GVD, propagation invariance is uniquely compatible with O-shaped wave packets \cite{Liou92PRA,Malaguti08OL,Dallaire09OE} that are circularly symmetric with respect to space and time \cite{Faccio07Book}, which are associated with closed \textit{elliptical} spatiotemporal spectra having an upper limit imposed on the utilizable bandwidth. Hence, tuning the wavelength through the zero-dispersion wavelength (ZDW) of the medium is expected to produce a \textit{gradual} X-to-O structural field transition as the GVD goes from normal to anomalous \cite{Porras05OL,Kandidov11JETP,Chekalin15JPB,Mitrofanov16OL,Ren19PRL,Panov21PRA}, with intermediate structures appearing when the spectrum straddles the ZDW \cite{Faccio07Book}.

\begin{figure*}[t!]
\centering
\includegraphics[width=17.6cm]{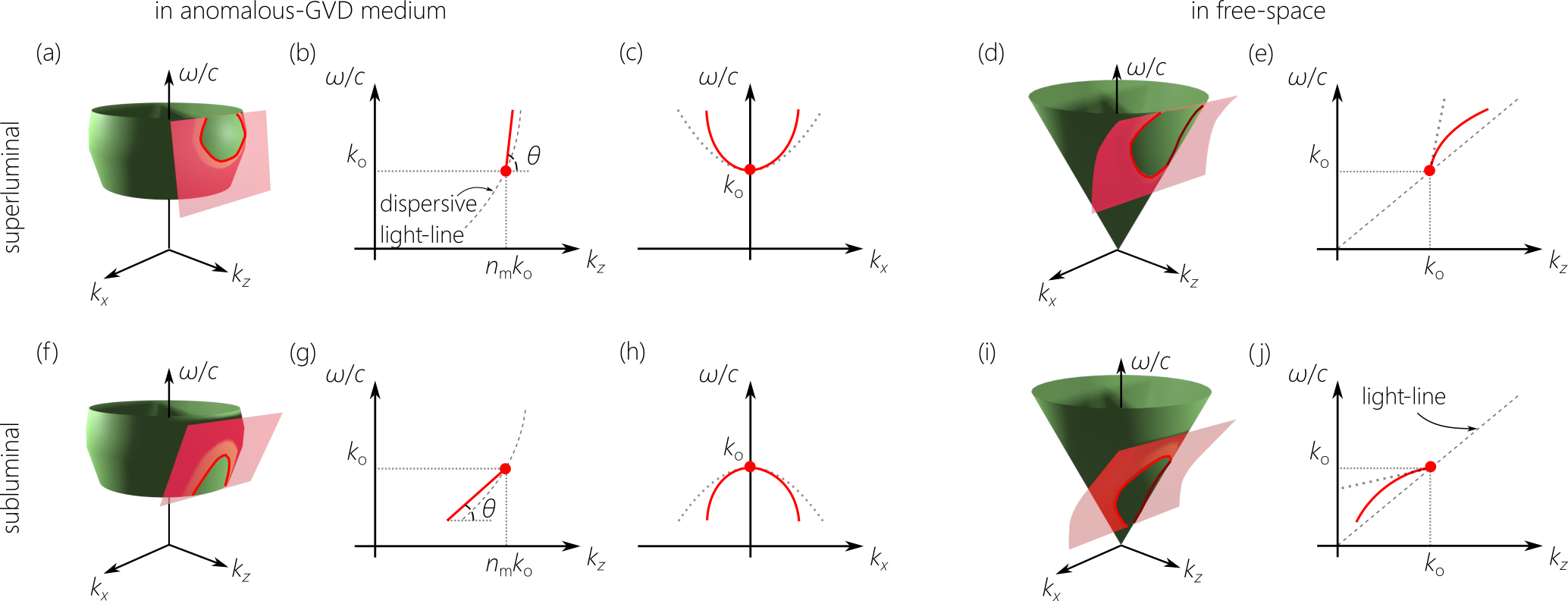}
\caption{(a) The spectral support for propagation-invariant STWPs in presence of anomalous GVD is the intersection of the dispersive light-cone $k_{x}^{2}+k_{z}^{2}\!=\!k^{2}$ with a tilted plane (Eq.~\ref{Eq:StraightLineProjection}). (b) The $(k_{z},\tfrac{\omega}{c})$-projection is a straight line, and (c) the $(k_{x},\tfrac{\omega}{c})$-projection is a conic section. The latter projection is invariant upon refraction at normal incidence onto a planar interface between free space and the dispersive medium. (d) The corresponding spectral support \textit{in free space} for the STWP in (a) is the intersection of the free-space light-cone $k_{x}^{2}+k_{z}^{2}\!=\!(\tfrac{\omega}{c})^{2}$ with a planar curved surface (Eq.~\ref{eq:kzFreeSpace}). (e) The $(k_{z},\tfrac{\omega}{c})$-projection is a curve consistent with \textit{normal} GVD in free space. Panels (a-e) correspond to a superluminal STWP $\widetilde{v}\!>\!\widetilde{v}_{\mathrm{m}}$, and (f-j) to a subluminal STWP $\widetilde{v}\!<\!\widetilde{v}_{\mathrm{m}}$.}
\label{Fig:Concept}
\end{figure*}

Previous experimental evidence for the existence of linear propagation-invariant wave packets in dispersive media has been gleaned from the light produced by certain nonlinear interactions, in which the phase-matching conditions in the nonlinear medium impose the required space-time coupling \cite{DiTrapani03PRL,Faccio06PRL,Faccio07OE,Porras07JOSAB}. The quest for observing the X-to-O transition has focused to date on only measuring the spatiotemporal \textit{spectrum}, which transitions gradually from a hyperbolic to an elliptical structure as the wavelength is tuned through the ZDW \cite{Porras05OL}, but the X-to-O transition in the field structure and the propagation invariance of the wave packets were not confirmed. Along similar lines, O-shaped spectra for entangled-photon pairs produced by spontaneous parametric downconversion have been recently observed in the anomalous-GVD regimes \cite{Spasibko16OL,Cutipa20OL}. Therefore, the X-to-O transition has yet to be observed directly. However, the theoretical study by Malaguti, Bellanca, and Trillo \cite{Malaguti08OL} (`MBT' henceforth) has pointed out a different strategy for realizing the X-to-O structural field transition \textit{at a fixed wavelength} in presence of anomalous GVD (see also the recent study in \cite{Bejot22ACSP}). By tuning the wave-packet group velocity and without changing the magnitude or sign of the GVD, it is predicted that the wave packet undergoes an \textit{abrupt} transition from an X-shaped to an O-shaped profile when the group velocity crosses a threshold value that we call the `escape velocity'. To the best of our knowledge, this fixed-wavelength, structural field transition has not been observed to date. 

The challenge of tuning the group velocity of a linear wave packet has been recently addressed through the development of `space-time wave packets' (STWPs) \cite{Kondakci16OE,Parker16OE,Yessenov22AOP}, which are propagation-invariant pulsed beams in linear media with readily tunable group velocity \cite{Salo01JOA,Wong17ACSP2,Porras17OL,Efremidis17OL,Kondakci19NC,Yessenov19OE,Bhaduri20NP}. This unique behavior \cite{Porras17OL,Efremidis17OL,Wong17ACSP2,Kondakci17NP} stems from introducing \textit{non-differentiable angular dispersion} \cite{Hall21OL,Hall21OL3NormalGVD,Hall22OEConsequences,Hall22JOSAA}: each temporal frequency $\omega$ travels at a different angle $\varphi(\omega)$ with respect to the propagation axis, in such a way that the derivative of $\varphi(\omega)$ with respect to $\omega$ is undefined at some frequency $\omega_{\mathrm{o}}$ \cite{Hall22OEConsequences}. Crucially, non-differentiable angular dispersion enables tuning the on-axis group-velocity of the STWP over a broad range of values while remaining in the paraxial regime. Varying the STWP group velocity in presence of anomalous GVD thus offers a path towards observing the predicted abrupt X-to-O transition.

Here we observe the predicted fixed-wavelength X-to-O structural field transition in the anomalous-GVD regime by synthesizing STWPs in free space that are endowed with the appropriate spatiotemporal spectral structure to propagate invariantly when coupled to the dispersive medium. Without changing the wavelength or the material GVD, this X-to-O transition occurs by tuning the spatiotemporal spectrum to vary the STWP group velocity. Propagation invariance in the dispersive medium imposes a linear relationship between the axial wave number and the temporal frequency, which determines the spatiotemporal spectral support for the STWP on the surface of the dispersive light-cone. Below a threshold group-velocity value, it is guaranteed that this linear constraint intersects twice with the curved light-line in the dispersive medium, thereby giving rise to a closed elliptical spatiotemporal spectrum and an O-shaped profile. Increasing the group velocity above this threshold value prevents re-intersection of the axial-wave-number constraint with the light-line, resulting in an open hyperbolic spatiotemporal spectrum and an X-shaped profile. We thus refer to this threshold value as the `escape velocity'.

Using 100-fs pulses at a wavelength $\approx\!1$~$\mu$m in presence of anomalous GVD, we synthesize STWPs that enable us to observe unambiguously the X-to-O transition in the spectral domain (from hyperbolic to elliptical spatiotemporal spectra) and in the physical space-time domain (from X-shaped to O-shaped spatiotemporal profiles) as the STWP group velocity is tuned through the escape-velocity threshold. We confirm the propagation invariance of these STWPs in the dispersive medium and their dispersive behavior in free space. Additionally, we examine the impact of reducing the bandwidth (or spectral truncation) on the formation of O-shaped STWPs with circular symmetry in space and time. These results pave the way to engineering new phase-matched interactions in nonlinear and quantum optics.

\section{Dispersion-free space-time wave packets in presence of anomalous GVD}

We first examine the spatiotemporal spectrum for a propagation-invariant STWP in presence of anomalous GVD. We expand the wave number in the medium $k(\omega)\!=\!n(\omega)\omega/c$ around the temporal frequency $\omega\!=\!\omega_{\mathrm{o}}$ to second-order in $\Omega\!=\!\omega-\omega_{\mathrm{o}}$: $k(\omega_{\mathrm{o}}+\Omega)\!\approx\! n_{\mathrm{m}}k_{\mathrm{o}}+\tfrac{\Omega}{\widetilde{v}_{\mathrm{m}}}-\tfrac{1}{2}|k_{2\mathrm{m}}|\Omega^{2}$; where $n(\omega)$ is the refractive index, $c$ is the speed of light in vacuum, $k_{\mathrm{o}}\!=\!\omega_{\mathrm{o}}/c$, $n_{\mathrm{m}}\!=\!n(\omega_{\mathrm{o}})$, the group velocity is $\widetilde{v}_{\mathrm{m}}\!=\!1\big/\tfrac{dk}{d\omega}\big|_{\omega_{\mathrm{o}}}$, the negative-valued GVD coefficient is $k_{2\mathrm{m}}\!=\!\tfrac{d^{2}k}{d\omega^{2}}\big|_{\omega_{\mathrm{o}}}\!\!=\!-|k_{2\mathrm{m}}|$ \cite{SalehBook07}, and we define a dimensionless GVD parameter $\beta\!=\!c\omega_{\mathrm{o}}|k_{2\mathrm{m}}|$. The bulk dispersion relationship $k_{x}^{2}+k_{z}^{2}\!=\!k^{2}$ is represented geometrically by a modified (dispersive) light-cone [Fig.~\ref{Fig:Concept}(a)]; here $k_{x}$ and $k_{z}$ are the transverse and longitudinal components of the wave vector along $x$ and $z$, respectively, and we hold the field uniform along $y$ for simplicity ($k_{y}\!=\!0$). 

Propagation invariance in this dispersive medium is ensured by enforcing the constraint \cite{Salo01JOA,Turunen10PO,FigueroaBook14,Yessenov19PRA}:
\begin{equation}\label{Eq:StraightLineProjection}
k_{z}(\Omega)=n_{\mathrm{m}}k_{\mathrm{o}}+\frac{\Omega}{\widetilde{v}};
\end{equation}
where $\widetilde{v}$ is the group velocity of the STWP in the medium. When $\widetilde{v}\!<\!\widetilde{v}_{\mathrm{m}}$ we refer to the STWP as subluminal, superluminal when $\widetilde{v}\!>\!\widetilde{v}_{\mathrm{m}}$, and luminal when $\widetilde{v}\!=\!\widetilde{v}_{\mathrm{m}}$. The constraint in Eq.~\ref{Eq:StraightLineProjection} corresponds geometrically to a plane that is parallel to the $k_{x}$-axis and makes an angle $\theta$ (the spectral tilt angle) with the $k_{z}$-axis, where $\widetilde{v}\!=\!c\tan{\theta}$. Hence, the group velocity $\widetilde{v}$ of the non-dispersive STWP in the dispersive medium can be tuned, in principle, independently of the medium properties by varying the internal degree of freedom $\theta$ \cite{Bhaduri19Optica,Bhaduri20NP,Motz21OL}.

The spectral support for this dispersion-free STWP is the curve at the intersection of the dispersive light-cone with the plane in Eq.~\ref{Eq:StraightLineProjection} [Fig.~\ref{Fig:Concept}(a) for the superluminal and Fig.~\ref{Fig:Concept}(f) for the subluminal cases]. The spectral projection onto the $(k_{z},\tfrac{\omega}{c})$-plane is a straight line, indicating that the STWP does not experience dispersion of any order in the medium, whereas the $(k_{x},\tfrac{\omega}{c})$-projection can be approximated by a conic section [Fig.~\ref{Fig:Concept}(a-c) for superluminal and Fig.~\ref{Fig:Concept}(f-h) for subluminal]:
\begin{equation}\label{Eq:GeneralEqForSTSpectrum}
\frac{k_{x}^{2}}{k_{\mathrm{o}}^{2}}\approx2\alpha_{1}\frac{\Omega}{\omega_{\mathrm{o}}}+\alpha_{2}\left(\frac{\Omega}{\omega_{\mathrm{o}}}\right)^{2},
\end{equation}
where we have ignored third- and higher-order terms in $\Omega$, and the dimensionless parameters $\alpha_{1}$ and $\alpha_{2}$ are:
\begin{equation}\label{Eq:alpha1alpha2}
\alpha_{1}=n_{\mathrm{m}}(\widetilde{n}_{\mathrm{m}}-\widetilde{n}),\;\;\;\alpha_{2}=\widetilde{n}_{\mathrm{e}}^{2}-\widetilde{n}^{2};
\end{equation}
here $\widetilde{n}_{\mathrm{m}}\!=\!c/\widetilde{v}_{\mathrm{m}}$ is the group index of the dispersive medium, $\widetilde{n}\!=\!c/\widetilde{v}\!=\!\cot{\theta}$ is that for the STWP in the medium, and $\widetilde{n}_{\mathrm{e}}\!=\!c/\widetilde{v}_{\mathrm{e}}$, where we call $\widetilde{v}_{\mathrm{e}}$ the `escape velocity' for reasons that will become clear shortly \cite{Malaguti08OL}:
\begin{equation}
\widetilde{v}_{\mathrm{e}}=\frac{\widetilde{v}_{\mathrm{m}}}{\sqrt{1-n_{\mathrm{m}}\beta\widetilde{v}_{\mathrm{m}}^{2}/c^{2}}}>\widetilde{v}_{\mathrm{m}},
\end{equation}
so that $\widetilde{n}_{\mathrm{e}}\!=\!\widetilde{n}_{\mathrm{m}}^{2}-n_{\mathrm{m}}\beta$.

However, such an STWP is first synthesized in free space, where the light-cone is $k_{x}^{2}+k_{z}^{2}\!=\!(\tfrac{\omega}{c})^{2}$ [Fig.~\ref{Fig:Concept}(d,i)], and is then coupled to the dispersive medium. The spectral support of the STWP on the free-space light-cone can be determined from its counterpart in the dispersive medium by taking advantage of the invariance of $k_{x}$ and $\omega$ upon refraction across a planar interface \cite{Bhaduri19Optica,Bhaduri20NP}. That is, the $(k_{x},\tfrac{\omega}{c})$-projection (Eq.~\ref{Eq:GeneralEqForSTSpectrum}) is invariant, so that the $(k_{z},\tfrac{\omega}{c})$-projection is no longer a straight line, and is instead curved [Fig.~\ref{Fig:Concept}(e,j)]:
\begin{equation}\label{eq:kzFreeSpace}
k_{z}\approx k_{\mathrm{o}}+(1-\alpha_{1})\frac{\Omega}{c}+\frac{1}{2}\left(\frac{1}{\omega_{\mathrm{o}}c}+n_{\mathrm{m}}|k_{2\mathrm{m}}|\Omega^{2}\right),
\end{equation}
where we have assumed that $\widetilde{n}\approx\widetilde{n}_{\mathrm{m}}$; that is, the spectral support of the STWP remains close th the light-line. Here $\tfrac{d^{2}k_{z}}{d\omega^{2}}\big|_{\omega_{\mathrm{o}}}\!=\!n_{\mathrm{m}}|k_{2\mathrm{m}}|\!>\!0$, indicating the presence of \textit{normal} GVD in free space that cancels the anomalous GVD in the medium, and the group velocity in free space $\widetilde{v}_{\mathrm{a}}$ is given by $\tfrac{1}{\widetilde{v}_{\mathrm{a}}}\!=\!\tfrac{dk_{z}}{d\omega}\big|_{\omega_{\mathrm{o}}}\!=\!\tfrac{1-\alpha_{1}}{c}$. Writing $\widetilde{n}_{\mathrm{a}}\!=\!c/\widetilde{v}_{\mathrm{a}}$ for the group index of the STWP in free space, we have $1-\widetilde{n}_{\mathrm{a}}\!=\!n_{m}(\widetilde{n}_{\mathrm{m}}-\widetilde{n})$, as shown in \cite{Bhaduri20NP,Yessenov22OLDispersiveRefraction,He21Arxiv}. Because $k_{x}$ and $\Omega$ are conserved at a planar interface between two media, both $\alpha_{1}$ and $\alpha_{2}$ (Eq.~\ref{Eq:GeneralEqForSTSpectrum}) are \textit{refractive invariants} for STWPs: $\alpha_{1}$ governs the change in group velocity of STWPs \cite{Bhaduri20NP,Motz21OL}, whereas $\alpha_{2}$ governs the change in GVD experienced by the STWP upon traversing the interface \cite{Yessenov22OLDispersiveRefraction,He21Arxiv}.

Equation~\ref{Eq:GeneralEqForSTSpectrum} indicates the presence of angular dispersion \cite{Torres10AOP} in the field structure; that is, each frequency $\omega$ travels at a different angle $\varphi(\omega)$ in free space with the $z$-axis, where $k_{x}(\omega)\!=\!\tfrac{\omega}{c}\sin\{\varphi(\omega)\}$. It is usually thought that angular dispersion can introduce only \textit{anomalous} GVD in free space, which can then be utilized for dispersion cancellation in a normal-GVD medium \cite{Martinez84JOSAA}. However, we have recently shown that the well-known result in \cite{Martinez84JOSAA} can be circumvented, and \textit{normal} GVD thus inculcated in free space to cancel the anomalous GVD in the medium, by introducing \textit{non-differentiable} angular dispersion \cite{Hall21OL,Hall21OL3NormalGVD,Hall22OEConsequences,Hall22JOSAA}; that is, there must exist at least one frequency at which $\varphi(\omega)$ does not possess a derivative. For example, the derivative of
$\varphi(\omega)\!\propto\!\sqrt{\omega-\omega_{\mathrm{o}}}$ is not defined at $\omega\!=\!\omega_{\mathrm{o}}$. It is clear that the AD expressed by Eq.~\ref{Eq:GeneralEqForSTSpectrum} is of this kind by examining the small-angle limit ($\sin\varphi\!\rightarrow\!\varphi$ as $\omega\!\rightarrow\!\omega_{\mathrm{o}}$) \cite{Hall21OL,Hall22JOSAA}.

\section{X-to-O Structural field transition}

\begin{figure*}[t!]
\centering
\includegraphics[width=18.4cm]{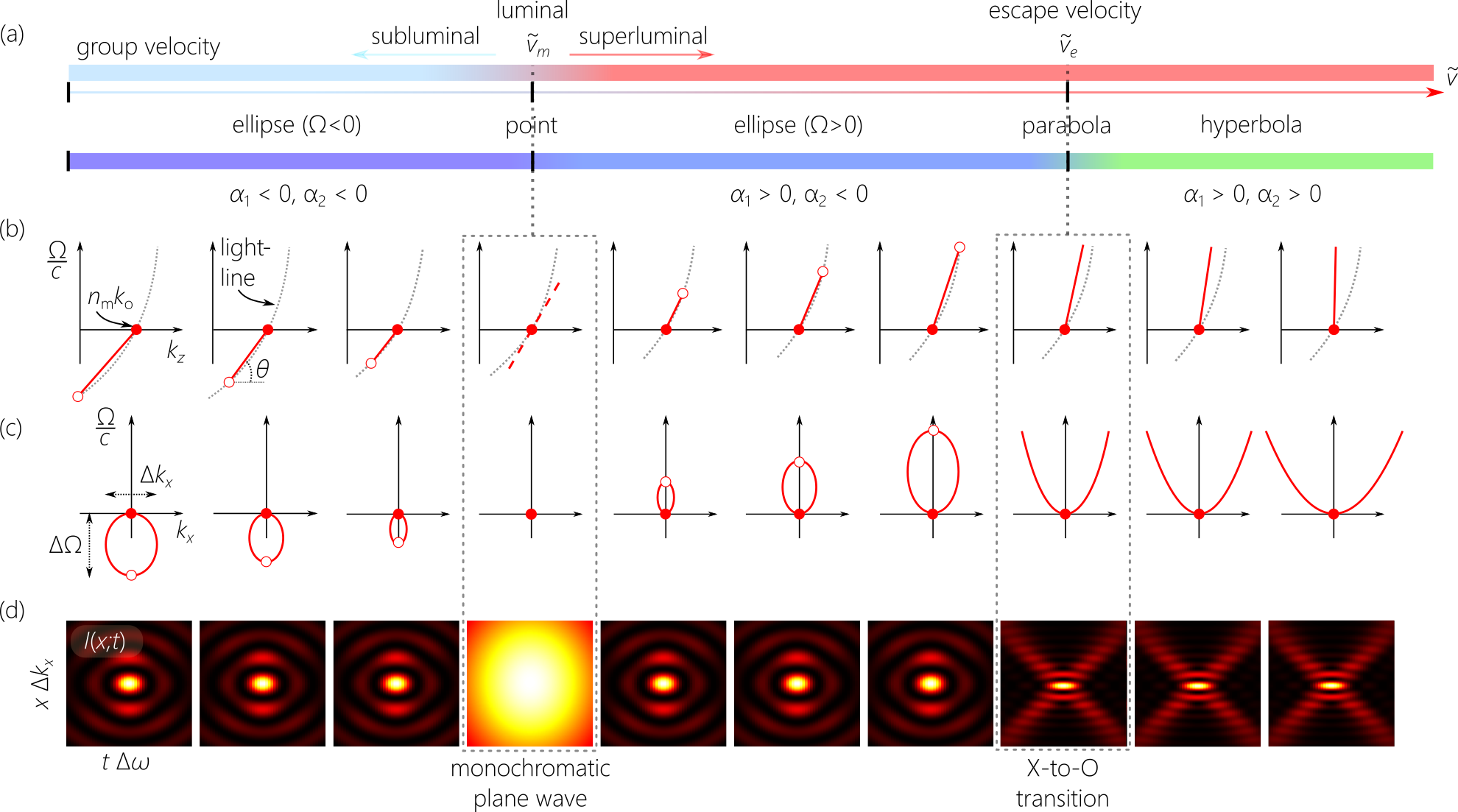}
\caption{(a) Classification of propagation-invariant STWPs in presence of anomalous GVD according to their group velocity $\widetilde{v}$. We highlight the two transition points at the luminal ($\widetilde{v}\!=\!\widetilde{v}_{\mathrm{m}}$), and the escape ($\widetilde{v}\!=\!\widetilde{v}_{\mathrm{e}}$) velocities. (b) The $(k_{z},\tfrac{\Omega}{c})$-projection and (c) the $(k_{x},\tfrac{\Omega}{c})$-projection while tuning $\widetilde{v}$. The solid circle at $\Omega\!=\!0$, corresponding to a fixed intersection point with the dispersive light-line at $(k_{x},k_{z},\tfrac{\omega}{c})\!=\!(0,n_{\mathrm{m}}k_{\mathrm{o}},k_{\mathrm{o}})$. A second intersection point (empty circle) exists only when $\alpha_{2}\!<\!0$, yielding an elliptical $(k_{x},\tfrac{\Omega}{c})$-projection. (d) Spatiotemporal intensity profiles $I(x;t)$ for the STWPs at a fixed $z$, showing the X-to-O transition at $\widetilde{v}\!=\!\widetilde{v}_{\mathrm{e}}$.}
\label{Fig:TransitionTheory}
\end{figure*}

The $(k_{x},\tfrac{\omega}{c})$-projection of the STWP in the dispersive medium (Eq.~\ref{Eq:GeneralEqForSTSpectrum}) corresponds to a conic section whose type depends on the signs of the two refractive invariants $\alpha_{1}$ and $\alpha_{2}$. Once $\omega_{\mathrm{o}}$ is selected and the material parameters ($n_{\mathrm{m}}$, $\widetilde{n}_{\mathrm{m}}$, and $k_{2\mathrm{m}}$) are fixed, tuning the group velocity $\widetilde{v}$ spans a one-parameter group of dispersion-free STWPs. Switching the signs of $\alpha_{1}$ and $\alpha_{2}$ as $\widetilde{v}$ changes is associated with \textit{abrupt} changes in the structure of the spatiotemporal spectrum \textit{and} the profile of the STWP. By varying the spectral tilt angle $\theta$ made with the $k_{z}$-axis by the plane in Eq.~\ref{Eq:StraightLineProjection} in the dispersive medium, two transitions are encountered [Fig.~\ref{Fig:TransitionTheory}(a)]: the first transition occurs at $\widetilde{v}\!=\!\widetilde{v}_{\mathrm{m}}$, which corresponds to the luminal condition, whereupon $\alpha_{1}\!=\!0$; and the second transition occurs at $\widetilde{v}\!=\!\widetilde{v}_{\mathrm{e}}$, which corresponds to the escape-velocity condition, whereupon $\alpha_{2}\!=\!0$. We proceed to describe the consequences of these two transition points.

\subsection{Classifying the field structures by the group velocity}

\subsubsection{Subluminal wave packet, elliptical spectrum \texorpdfstring{$(\widetilde{v}\!<\!\widetilde{v}_{\mathrm{m}})$}{TEXT}}

We start off in the subluminal regime $\widetilde{v}\!<\!\widetilde{v}_{\mathrm{m}}$ in the dispersive medium, where $\alpha_{1},\alpha_{2}\!<\!0$. The $(k_{z},\tfrac{\Omega}{c})$-projection is a straight line that makes an angle $\theta$ with the $k_{z}$-axis, always passing through the point $(k_{z},\tfrac{\Omega}{c})\!=\!(n_{\mathrm{m}}k_{\mathrm{o}},0)$ on the dispersive light-line. The existence of a second intersection point with the light-line depends on $\theta$ [Fig.~\ref{Fig:TransitionTheory}(b)]. When $\widetilde{v}\!<\!\widetilde{v}_{\mathrm{m}}$, the $(k_{z},\tfrac{\Omega}{c})$-projection re-intersects with the dispersive light-line at $\tfrac{\Omega_{-}}{\omega_{\mathrm{o}}}\!=\!-\tfrac{2}{\beta}|\widetilde{n}_{\mathrm{m}}-\widetilde{n}|\!<\!0$, and the $(k_{x},\tfrac{\omega}{c})$-projection is an ellipse in the domain $\Omega\!<\!0$ [Fig.~\ref{Fig:TransitionTheory}(c)]:
\begin{equation}\label{eq:Ellipse}
\left(\frac{\Omega}{\Omega_{\mathrm{ST}}}+1\right)^{2}+\left(\frac{k_{x}}{k_{\mathrm{ST}}}\right)^{2}=1;
\end{equation}
here $\Omega_{\mathrm{ST}}/\omega_{\mathrm{o}}\!=\!|\alpha_{1}/\alpha_{2}|$ and $k_{\mathrm{ST}}/k_{\mathrm{o}}\!=\!|\alpha_{1}|/\sqrt{|\alpha_{2}|}$. A spatiotemporal spectrum in the form of a closed ellipse results in an O-shaped spatiotemporal profile $I(x;\tau)$ that is circularly symmetric in space and time \cite{Malaguti08OL,Dallaire09OE,Hall23NP,Diouf23NP}.

The temporal and spatial bandwidths have upper limits $\Delta\omega\!=\!|\Omega_{-}|$ and $\Delta k_{x}\!\sim\!k_{\mathrm{ST}}$, respectively, determined by the size of this ellipse. A larger bandwidth is \textit{not} compatible with propagation invariance at this group velocity in the dispersive medium. Lowering $\widetilde{v}$ increases the maximum achievable temporal and spatial bandwidths, resulting in more spatiotemporally localized profiles. On the other hand, when approaching the luminal limit $\widetilde{v}\!\rightarrow\!\widetilde{v}_{\mathrm{m}}$, the $(k_{z},\tfrac{\Omega}{c})$-projection becomes tangential to the dispersive light-line at $\Omega\!=\!0$, and the elliptical spectrum in the $(k_{x},\tfrac{\Omega}{c})$-projection collapses to a point. This indicates that the only \textit{luminal} field compatible with propagation invariance in presence of anomalous GVD is a monochromatic plane wave. Indeed, when $\widetilde{v}\!=\!\widetilde{v}_{\mathrm{m}}$, we have $\alpha_{1}\!=\!0$, while $\alpha_{2}$ remains negative-valued, so that $k_{x}/k_{\mathrm{o}}\!=\!\pm i\sqrt{|\alpha_{2}|}\Omega/\omega_{\mathrm{o}}$, which corresponds to evanescent waves except when $\Omega\!=\!0$.

\subsubsection{Superluminal wave packet, elliptical spectrum \texorpdfstring{$(\widetilde{v}_{\mathrm{m}}\!<\!\widetilde{v}\!<\!\widetilde{v}_{\mathrm{e}})$}{TEXT}}

When $\widetilde{v}\!>\!\widetilde{v}_{\mathrm{m}}$, we enter the superluminal regime. In the range $\widetilde{v}_{\mathrm{m}}\!<\!\widetilde{v}\!<\!\widetilde{v}_{\mathrm{e}}$ limited from above by the escape velocity, we have $\alpha_{1}\!>\!0$ but $\alpha_{2}\!<\!0$, and the spectral line in the $(k_{z},\tfrac{\Omega}{c})$-projection re-intersects with the dispersive light-line at $\tfrac{\Omega_{+}}{\omega_{\mathrm{o}}}\!=\!\tfrac{2}{\beta}(\widetilde{n}_{\mathrm{m}}-\widetilde{n})\!>\!0$. In the $(k_{x},\tfrac{\Omega}{c})$-projection, the spatiotemporal spectrum remains a closed ellipse in the domain $\Omega\!>\!0$:
\begin{equation}
\left(\frac{\Omega}{\Omega_{\mathrm{ST}}}-1\right)^{2}+\left(\frac{k_{x}}{k_{\mathrm{ST}}}\right)^{2}=1;
\end{equation}
and the spatiotemporal profile therefore remains O-shaped, with maximum temporal and spatial bandwidths $\Delta\omega\!=\!\Omega_{+}$ and $\Delta k_{x}\!\sim\!k_{\mathrm{ST}}$, respectively. Increasing $\widetilde{v}$ above the luminal limit $\widetilde{v}_{\mathrm{m}}$ increases the utilizable spectrum. 

When the escape velocity is reached $\widetilde{v}\!=\!\widetilde{v}_{\mathrm{e}}$, we have $\alpha_{2}\!=\!0$ and $\alpha_{1}\!>\!0$, and the $(k_{x},\tfrac{\Omega}{c})$-projection becomes abruptly an open \textit{parabola} in the domain $\Omega\!>\!0$ rather than a closed ellipse:
\begin{equation}
\frac{k_{x}^{2}}{k_{\mathrm{o}}^{2}}\!=\!2\alpha_{1}\frac{\Omega}{\omega_{\mathrm{o}}}.
\end{equation}
The escape velocity $\widetilde{v}_{\mathrm{e}}$ is thus the smallest group velocity (or, equivalently, the smallest spectral tilt angle $\theta$) for which the spectral line in the $(k_{z},\tfrac{\Omega}{c})$-projection escapes from re-intersection with the light-line; hence the name `escape velocity'. Once $\widetilde{v}$ reaches $\widetilde{v}_{\mathrm{e}}$, the propagation-invariant spatiotemporal profile $I(x;t)$ becomes abruptly X-shaped, whose spatial and temporal bandwidths can in principle be extended indefinitely.

\subsubsection{Superluminal wave packet, hyperbolic spectrum \texorpdfstring{$(\widetilde{v}\!>\!\widetilde{v}_{\mathrm{e}})$}{TEXT}}

In the superluminal regime with $\widetilde{v}\!>\!\widetilde{v}_{\mathrm{e}}$, we have $\alpha_{1}\!>\!0$ and $\alpha_{2}\!>\!0$, in which case the spectral line does \textit{not} re-intersect with the light-line, and the $(k_{x},\tfrac{\omega}{c})$-projection is a hyperbola for $\Omega\!>\!0$:
\begin{equation}
\left(\frac{\Omega}{\Omega_{\mathrm{ST}}}+1\right)^{2}-\left(\frac{k_{x}}{k_{\mathrm{ST}}}\right)^{2}=1,
\end{equation}
whereupon the spatial and temporal bandwidths can be increased, in principle, without bound. The spatiotemporal profile $I(x;t)$ of the propagation-invariant STWP is X-shaped.

The field has thus undergone a structural transition: when $\widetilde{v}\!\geq\!\widetilde{v}_{\mathrm{e}}$ propagation invariance is consistent with only an X-shaped profile having a hyperbolic or parabolic spectrum with no upper limit on the bandwidth; whereas propagation invariance below $\widetilde{v}_{\mathrm{e}}$ is consistent with an O-shaped field having an elliptical spectrum with finite allowable bandwidth. We refer to this abrupt change in the field configuration at $\widetilde{v}\!=\widetilde{v}_{\mathrm{e}}$ as an X-to-O transition. We emphasize that this transition occurs at a fixed wavelength, and thus the magnitude and sign of the (anomalous) GVD encountered by the field in the medium has not changed -- only the STWP group velocity $\widetilde{v}$ was tuned through $\widetilde{v}_{\mathrm{e}}$.

\section{Experimental configuration}

The dispersive sample we utilize comprises a pair of chirped Bragg mirrors (Edmund 12-335) that provide anomalous group-delay dispersion (GDD). By adjusting the separation between the mirrors and the incident angle of the STWP, we can control the number of bounces encountered by the STWP off the mirrors. Because the thickness of the mirrors is negligible with respect to the free-space gap between them, we take $n_{\mathrm{m}}\!\approx\!1$ and $\widetilde{v}_{\mathrm{m}}\!\approx\!c$ ($\widetilde{n}_{\mathrm{m}}\!\approx\!1$). The GVD coefficient $k_{2\mathrm{m}}\!\approx\!-500$~fs$^{2}$/mm is obtained by dividing the accumulated GDD by the propagation distance through the device, corresponding to a dimensionless GVD parameter $\beta\!\approx\!-0.25$, and an escape velocity of $\widetilde{v}_{\mathrm{e}}\!\approx\!1.17c$.

\begin{figure}[t!]
\centering
\includegraphics[width=8.6cm]{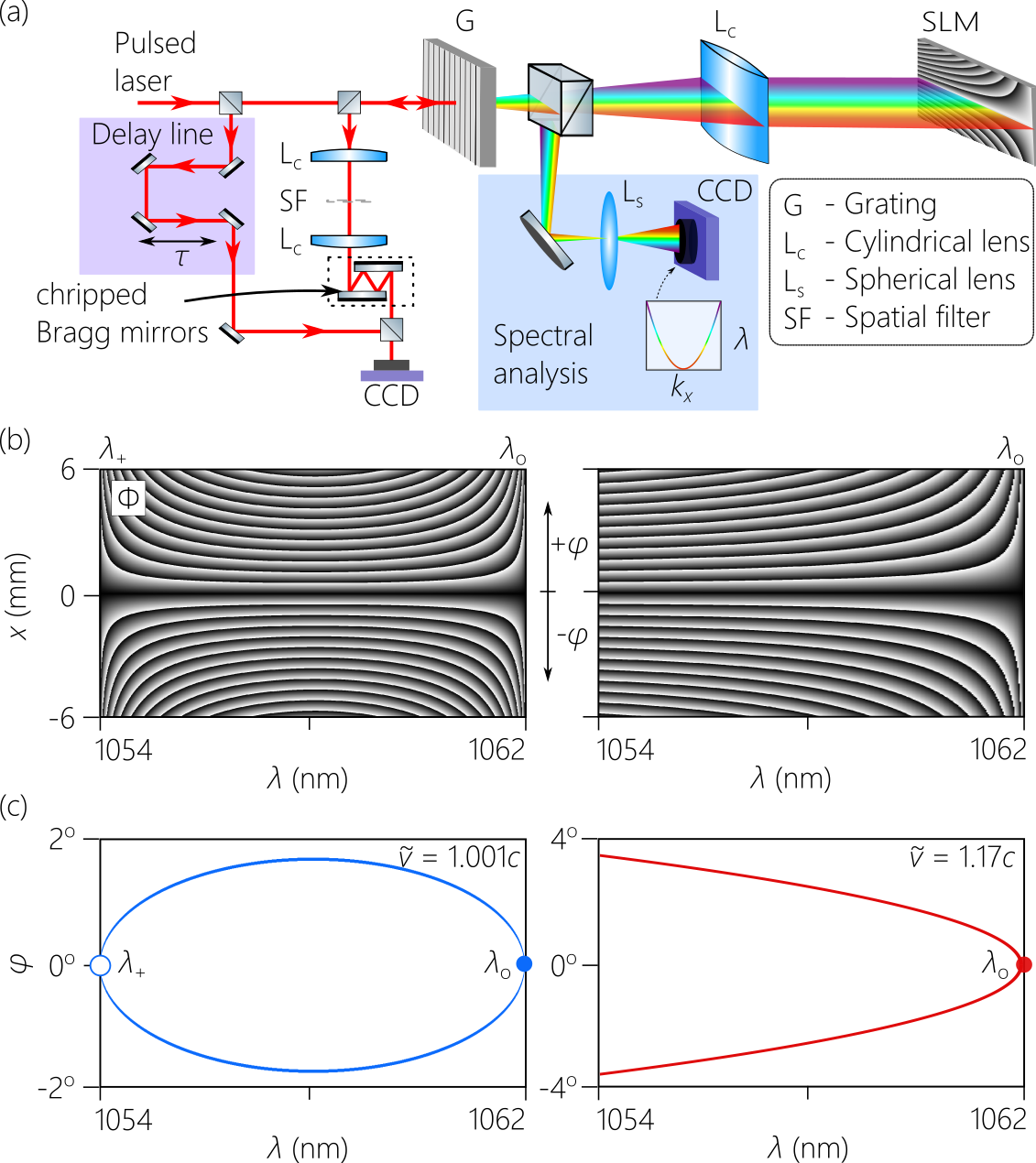}
\caption{(a) Setup for synthesizing and characterizing dispersive STWPs in free space. (b) Phase patterns imparted by the SLM to the spectrally resolved wave front for superluminal STWPs: $\widetilde{v}_{\mathrm{m}}\!<\!\widetilde{v}\!<\!\widetilde{v}_{\mathrm{e}}$ and $\widetilde{v}\!>\!\widetilde{v}_{\mathrm{e}}$. (c) The propagation angle $\varphi(\lambda)$ associated with the phase distributions in (b).}
\label{Fig:Setup}
\end{figure}

To produce non-dispersive STWPs in a dispersive medium, we first synthesize STWPs that are appropriately dispersive in free space. This necessitates precise control over the angular dispersion introduced into a generic pulsed beam by associating each frequency $\omega$ with a prescribed propagation angle $\varphi(\omega)$ with respect to the $z$-axis, with $\varphi(\omega_{\mathrm{o}})\!=\!0$. This is achieved using the universal angular-dispersion synthesizer depicted in Fig.~\ref{Fig:Setup}(a), which is based on our previous work in \cite{Yessenov22AOP,Hall23Normal}. Starting with femtosecond pulses from a mode-locked laser (Spark Lasers; Alcor) centered at a wavelength $\lambda\!=\!1064$~nm, of bandwidth $\Delta\lambda\!\approx\!20$~nm, and pulsewidth $\Delta T\!\approx\!100$~fs, the pulse spectrum is resolved spatially using a grating (1200~lines/mm) and the $-1$~diffraction order is collimated with a cylindrical lens (focal length $f\!=\!500$~mm). A reflective, phase-only spatial light modulator (SLM; Meadowlark, E19X12) placed at the focal plane deflects each wavelength $\lambda$ at angles $\pm\varphi(\lambda)$, where:
\begin{equation}
\sin\{\varphi(\Omega)\}\approx\frac{\pm1}{1+\frac{\Omega}{\omega_{\mathrm{o}}}}\sqrt{2\alpha_{1}\frac{\Omega}{\omega_{\mathrm{o}}}+\alpha_{2}\left(\frac{\Omega}{\omega_{\mathrm{o}}}\right)^{2}}.
\end{equation}
The retro-reflected wave front traces its path back through the cylindrical lens to the grating, where the STWP is formed.

Two examples of the phase distributions $\Phi$ imparted by the SLM are shown in Fig.~\ref{Fig:Setup}(b) for an O-shaped profile with $\widetilde{v}\!=\!1.001c$ and an X-shaped profile with $\widetilde{v}\!=\!1.17c$, and we plot in Fig.~\ref{Fig:Setup}(c) the corresponding angular-dispersion spectrum $\varphi(\lambda)$. In the former example, $\varphi\!=\!0$ (flat phase $\Phi\!=\!0$) at $\lambda_{\mathrm{o}}\!=\!1062$~nm and $\lambda_{-}\!=\!1054$~nm in the closed elliptical spectrum (first and last columns of the SLM phase pattern), both associated with non-differential angular dispersion; whereas $\varphi\!=\!0$ in the latter example occurs only at $\lambda_{\mathrm{o}}\!=\!1062$~nm.

\begin{figure*}[t!]
\centering
\includegraphics[width=13.6cm]{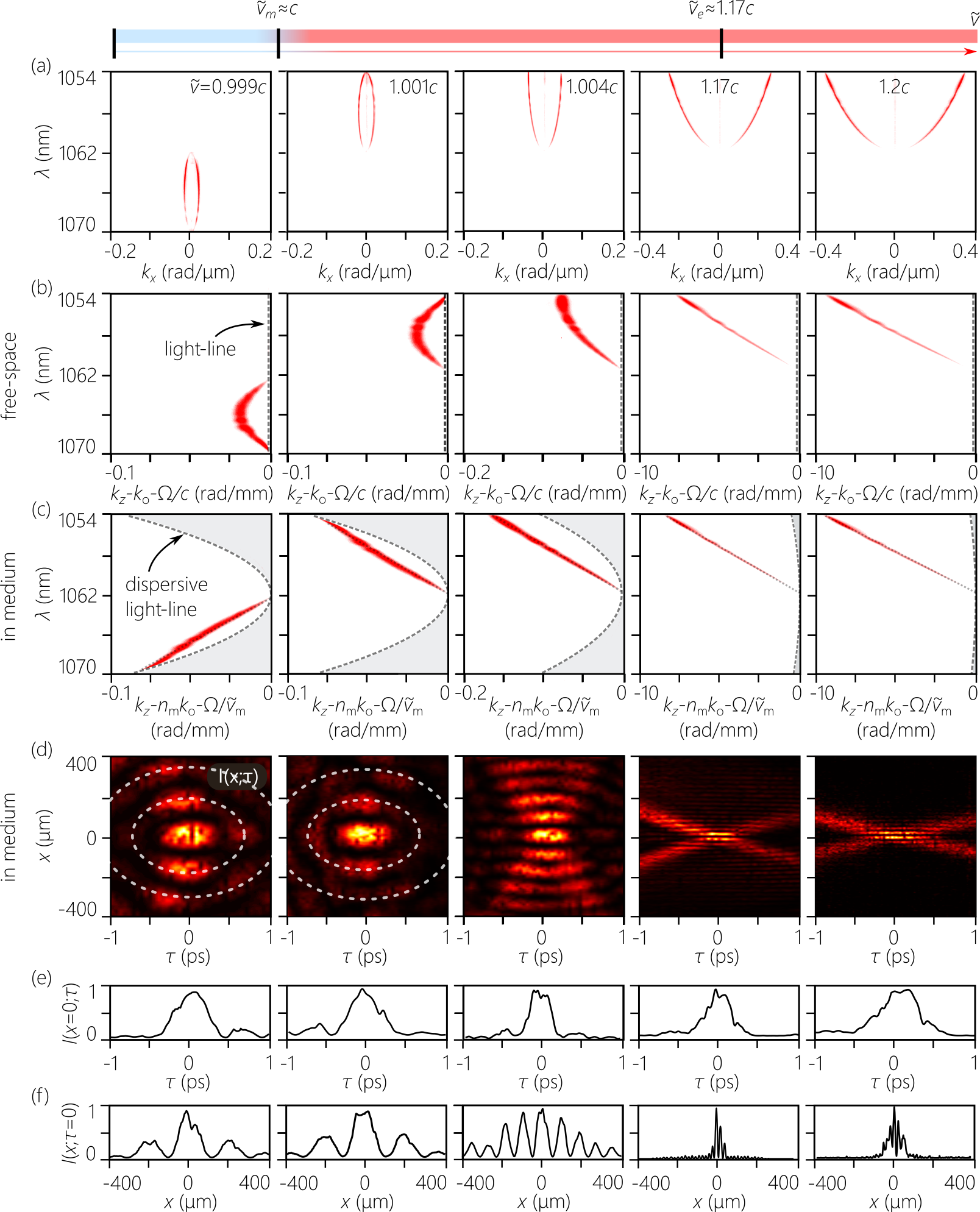}
\caption{(a) Measured spectral $(k_{x},\lambda)$-projections as $\widetilde{v}$ in the medium increases from left to right. (b) The $(k_{z},\lambda)$-projections in free space and (c) in the medium, extracted from (a). Here, $n_{\mathrm{m}}\!\approx\!1$ and $\widetilde{v}_{\mathrm{m}}\!\approx\!c$, so the horizontal scales in (b) and (c) are identical. Note however the change in the horizontal scale starting at $\widetilde{v}\!=\!\widetilde{v}_{\mathrm{e}}$. In (a-c), the wavelength $\lambda$ is plotted in reverse order to match the plots against $\Omega$ in Fig.~\ref{Fig:TransitionTheory}(c,d). (d) Measured spatiotemporal intensity profiles $I(x;\tau)$ at $z\!=\!0$ corresponding to the spectra in (a). (e) The temporal pulse profile $I(x\!=\!0;\tau)$ at the beam center $x\!=\!0$ obtained from (d). Because the temporal bandwidth $\Delta\lambda$ is held fixed, this temporal profile does not change with $\widetilde{v}$. (f) The spatial beam profile $I(x;\tau\!=\!0)$ at the pulse center $\tau\!=\!0$ obtained from (d).}
\label{Fig:DataScanningV}
\end{figure*}

The synthesized spatiotemporal spectrum is measured by implementing a spatial Fourier transform on the spectrally resolved wave front retro-reflected from the SLM, thereby producing the $(k_{x},\lambda)$-projection. We can then extract from this measurement the $(k_{z},\lambda)$-projection: in free space $k_{z}(\lambda)\!=\!\sqrt{(\tfrac{\omega}{c})^{2}-k_{x}^{2}}$ [Fig.~\ref{Fig:Concept}(e,j)], and in the dispersive medium [Fig.~\ref{Fig:Concept}(b,g)]:
\begin{equation}
k_{z}(\lambda)=\sqrt{\left(n_{\mathrm{m}}k_{\mathrm{o}}+\frac{\Omega}{\widetilde{v}_{\mathrm{m}}}-\frac{1}{2}|k_{2\mathrm{m}}|\Omega^{2}\right)^{2}-k_{x}^{2}}.
\end{equation}
The latter must be linear in $\omega$ if dispersion-free propagation in the medium is to be achieved (corresponding to Eq.~\ref{Eq:StraightLineProjection}).

To reconstruct the spatiotemporal intensity of the STWP envelope $I(x,z;\tau)$ at a fixed axial plane $z$, we separate a portion of the initial plane-wave laser pulses from the source prior to the grating and direct it to an optical delay line $\tau$ [Fig.~\ref{Fig:Setup}(a)]. We bring the synthesized STWP together with this reference pulse at a beam splitter, and adjust $\tau$ to overlap the two wave packets in space and time at a CCD camera placed at $z$. Spatially resolved interference fringes are observed, and their visibility is used to reconstruct $I(x,z;\tau)$ at this axial plane and at this relative delay \cite{Kondakci19NC,Yessenov19OE}. Sweeping $\tau$ thus enables reconstruction of the STWP spatiotemporal intensity profile at this fixed axial plane. 

\begin{figure*}[t!]
\centering
\includegraphics[width=17.6cm]{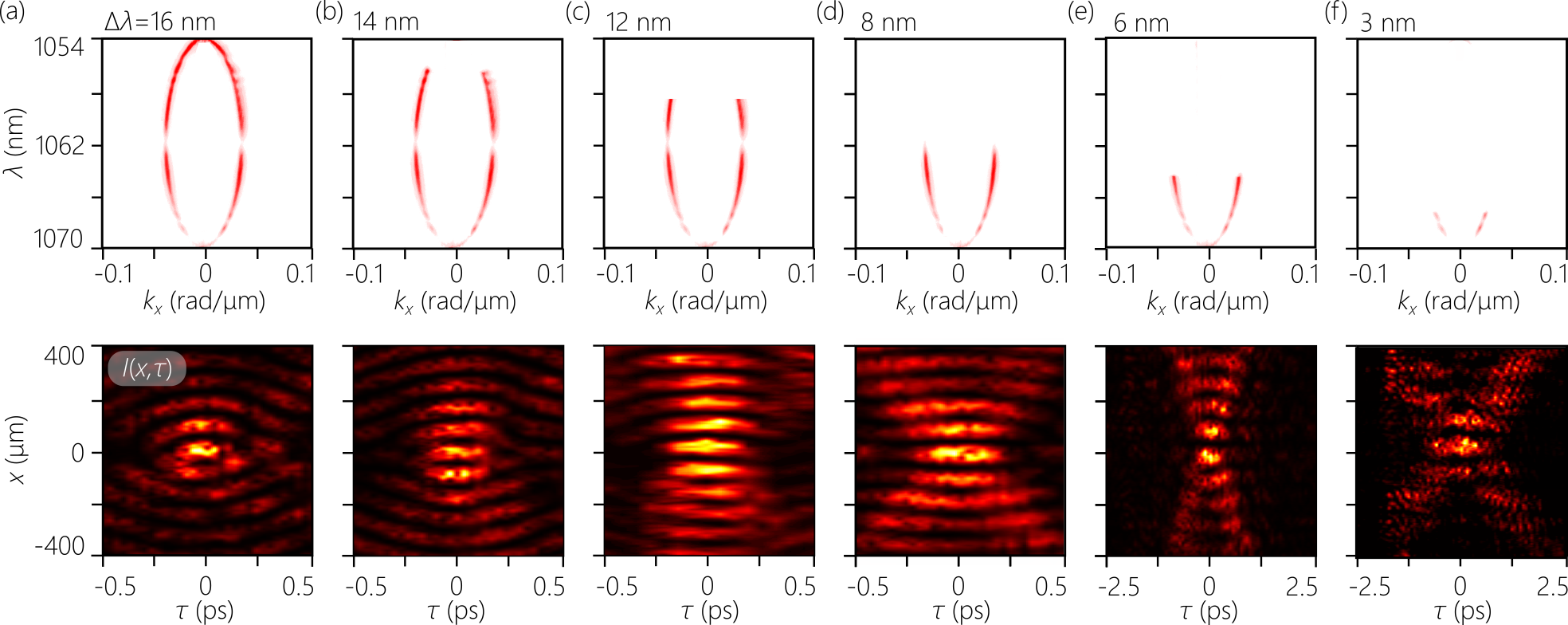}
\caption{Transition from an O-shaped to an X-shaped STWP in presence of anomalous GVD by changing the bandwidth $\Delta\lambda$. In the first row, we plot the measured spectral $(k_{x},\lambda)$-projections, and in the second row the reconstructed intensity profiles $I(x;\tau)$ at $z\!=\!30$~mm. From left to right we decrease $\Delta\lambda$ at $\widetilde{v}\!=\!1.004c$: (a) $\Delta\lambda\!\approx\!16$~nm, (b) 14~nm, (c) 12~nm, (d) 8~nm, (e) 6~nm, and (f) 3~nm}
\label{Fig:OWaveTransition}
\end{figure*}

\section{X-to-O transition by group-velocity tuning}

The measured $(k_{x},\lambda)$-projections are plotted in Fig.~\ref{Fig:DataScanningV}(a) as we increase $\widetilde{v}$ from $0.999c$ to $1.2c$, while maintaining the bandwidth at $\Delta\lambda\!\approx\!8$~nm with $\lambda_{\mathrm{o}}\!\approx\!1062$~nm. We proceed from a subluminal STWP at $\widetilde{v}\!=\!0.999c\!<\!\widetilde{v}_{\mathrm{m}}$ (Eq.~\ref{eq:Ellipse}) with a closed elliptical spectrum extending from $\lambda_{\mathrm{o}}\!=\!1062$~nm to $\lambda_{\mathrm{-}}\!=\!1070$~nm ($\Omega\!<\!0$). Increasing $\widetilde{v}$ past $\widetilde{v}_{\mathrm{m}}$ into the superluminal regime ($\widetilde{v}_{\mathrm{m}}\!<\!\widetilde{v}\!<\!\widetilde{v}_{\mathrm{e}}$) results in elliptical spectra flipped around $\lambda_{\mathrm{o}}$ with respect to their subluminal counterpart. When $\widetilde{v}\!=\!1.001c$, the spectrum is a closed ellipse extending from $\lambda_{\mathrm{o}}\!=\!1062$~nm to $\lambda_{+}\!=\!1054$~nm ($\Omega\!>\!0$). Increasing the group velocity to $\widetilde{v}\!=\!1.004c$ results in an increase in the maximum utilizable spatial and temporal bandwidths for the elliptical spectrum. However, we maintain the temporal bandwidth fixed at $\Delta\lambda\!=\!8$~nm, so that the elliptical spectrum is truncated. At the escape velocity $\widetilde{v}\!=\!\widetilde{v}_{\mathrm{e}}\!=\!1.17c$, the spectrum is no longer closed, and instead takes the form of a parabola. Finally, at $\widetilde{v}\!=\!1.2c\!>\!\widetilde{v}_{\mathrm{e}}$, the spectrum takes the form of a hyperbola. We extract from these spectral projections $(k_{z},\lambda)$-projections in free space [Fig.~\ref{Fig:DataScanningV}(b)] and in the dispersive medium [Fig.~\ref{Fig:DataScanningV}(c)]. In the former, the spectral projection is curved, indicating that the STWP experiences normal GVD in free space. In the latter, the spectral projections are straight lines, indicating non-dispersive propagation in the medium.

We plot in Fig.~\ref{Fig:DataScanningV}(d) the measured spatiotemporal intensity profiles of the STWPs at $z\!=\!0$. Whenever the $(k_{x},\lambda)$-projection is a closed trajectory, corresponding to the $(k_{z},\lambda)$-projection intersecting with the light-line \textit{twice}, the profile is O-shaped. Consequently, both the subluminal STWP at $\widetilde{v}\!=\!0.999c$ and the superluminal STWP at $\widetilde{v}\!=\!1.001c$ have O-shaped profiles $I(x;\tau)$ [Fig.~\ref{Fig:DataScanningV}(d)]. Because the spatial and temporal bandwidths are the same for both STWPs, the temporal pulse profile at the beam center $I(0;\tau)$ [Fig.~\ref{Fig:DataScanningV}(e)] and the spatial beam profile at the pulse center $I(x;0)$ [Fig.~\ref{Fig:DataScanningV}(f)] are also the same for both STWPs. However, because the elliptical spectrum is not closed at $\widetilde{v}\!=\!1.004c$, the profile is no longer O-shaped, the circular symmetry in space and time is broken, and further sidelobes are observed along $x$ [Fig.~\ref{Fig:DataScanningV}(d)]. The pulse profile $I(0;\tau)$ is unaffected by this truncation (with $\Delta\lambda$ held fixed) [Fig.~\ref{Fig:DataScanningV}(e)]; however, the increased spatial bandwidth results in a narrower central spatial peak in $I(x;0)$ [Fig.~\ref{Fig:DataScanningV}(f)].    

When the $(k_{x},\lambda)$-projections are parabolic or hyperbolic, and thus intrinsically open, the spatiotemporal profiles are X-shaped as is usual for STWPs in the paraxial regime. Once again, the pulse profiles at the beam center $I(0;\tau)$ are unaffected by the change in $\widetilde{v}$ [Fig.~\ref{Fig:DataScanningV}(e)]. However, the increased spatial bandwidth accompanying $\widetilde{v}\!\geq\!\widetilde{v}_{\mathrm{e}}$ results in much narrower beam profiles at the pulse center $I(x;0)$ [Fig.~\ref{Fig:DataScanningV}(f)]. We thus observe clearly the X-to-O structural field transition in Fig.~\ref{Fig:DataScanningV}(d) as $\widetilde{v}$ is tuned through $\widetilde{v}_{\mathrm{e}}$.

\section{Impact of varying the bandwidth}

Observing the O-shaped STWP profile in presence of anomalous GVD requires exploiting the full ellipse in the $(k_{x},\lambda)$-projection. Spectral truncation prevents the formation of the O-shaped profile as shown in Fig.~\ref{Fig:OWaveTransition} for the STWP with $\widetilde{v}\!=\!1.004c$. The maximum exploitable bandwidth in this case is $\Delta\lambda\!\approx\!16$~nm, at which point an O-shaped profile emerges [Fig.~\ref{Fig:OWaveTransition}(a)]. We gradually truncate the spectrum (the ellipse is \textit{not} closed), whereupon the O-shaped profile is only partially formed, and the concentric circular fringes are no longer closed [Fig.~\ref{Fig:OWaveTransition}(b,c)]. Further reducing the bandwidth to the midway point $\Delta\lambda\!\approx\!8$~nm [Fig.~\ref{Fig:OWaveTransition}(d)] results in the gradual emergence of an X-shaped profile, which becomes clearer as $\Delta\lambda$ is subsequently reduced [Fig.~\ref{Fig:OWaveTransition}(e)]. For small bandwidths, whereupon the spectrum is well-approximated by a parabola centered at $k_{x}\!=\!0$, the standard X-shaped profile for STWPs becomes clear [Fig.~\ref{Fig:OWaveTransition}(f)]. These results emphasize the necessity of exploiting the full available bandwidth in the formation of the O-shaped profile. Our recent work on propagation-invariant STWPs in dispersive media failed to observe the X-to-O-wave transition because the bandwidth utilized was not sufficient for such a distinction to emerge \cite{Hall23Normal}.

\begin{figure*}[t!]
\centering
\includegraphics[width=17.6cm]{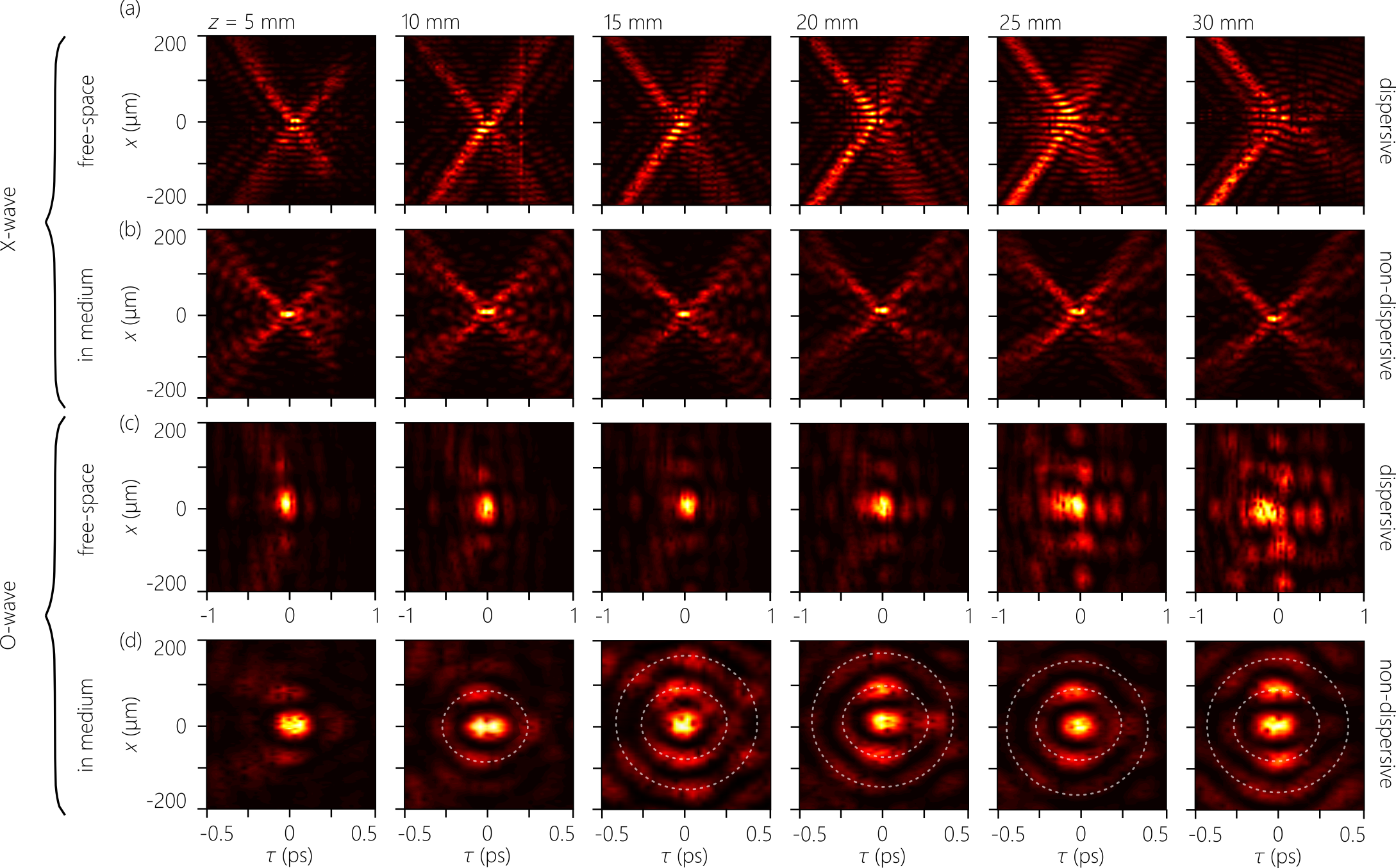}
\caption{(a) Spatiotemporal intensity profiles $I(x,z;\tau)$ for an X-shaped STWP reconstructed at 5-mm intervals along $z$ in free space ($\widetilde{v}_{\mathrm{a}}\!\approx\!1.04c$) showing temporal broadening consistent with normal GVD ($c\omega_{\mathrm{o}}k_{2\mathrm{a}}\!=\!0.25$). (b) Intensity profiles for the dispersive-free STWP in the dispersive medium ($\widetilde{v}\!=\!1.04c$). (c) Intensity profiles for an O-shaped STWP ($\widetilde{v}_{\mathrm{a}}\!\approx\!1.004c$) in free space. The STWP appears to split into multiple dispersive wave packets of different group velocities; see Fig.~\ref{Fig:PropagationInvariance2}. (d) Intensity profiles for the dispersion-free STWP in the dispersive medium ($\widetilde{v}\!=\!1.004c$). All profiles are acquired in a reference frame moving at $c$ ($\tau\!=\!t-z/c$). For clarity, we changed the horizontal scale for $\tau$ in (d) compared to (a-c).}
\label{Fig:PropagationInvariance}
\end{figure*}

\section{Propagation invariance in presence of GVD}

It is critical to keep in mind that this X-to-O transition is observed while tuning $\widetilde{v}$ at fixed $\lambda_{\mathrm{o}}$, so the GVD in the medium remains anomalous and of the same magnitude. Moreover, all these STWPs across this structural transition are propagation-invariant in the dispersive medium, although they are dispersive in free space. We verify these characteristics in Fig.~\ref{Fig:PropagationInvariance} by plotting the measured spatiotemporal profiles of X- and O-shaped STWPs measured at 5-mm axial intervals in both free space and in the dispersive medium. We plot in Fig.~\ref{Fig:PropagationInvariance}(a) the profiles for an X-shaped dispersive STWP in free space corresponding to a propagation-invariant STWP traveling at $\widetilde{v}_{\mathrm{a}}\!=\!1.04c$, where the STWP is dispersive and undergoes temporal broadening with propagation distance. With a temporal bandwidth of $\Delta\lambda\!\approx\!16$~nm, the pulsewidth at the beam center is 120~fs, and the dispersion length is $\approx\!12$~mm, leading to an increase of pulsewidth to 500~fs after 30~mm. The same STWP upon coupling to the medium is dispersion-free [Fig.~\ref{Fig:PropagationInvariance}(b)]. Similar results are obtained for the O-shaped STWP traveling at $\widetilde{v}\!=\!1.004c$ in the medium. In free space the STWP is dispersive [Fig.~\ref{Fig:PropagationInvariance}(c)], but is free of dispersion in the medium [Fig.~\ref{Fig:PropagationInvariance}(d)].

\section{Discussion and Conclusion}

Why has it been so challenging to synthesize linear non-dispersive O-shaped STWPs? Consider first the free-space scenario \cite{Yessenov22AOP}. The closed elliptical spatiotemporal spectrum required to produce an O-shaped profile can be realized in free space only at the intersection of the free-space light-cone with a plane when $\theta\!<\!45^{\circ}$ and retaining \textit{all} values of $k_{z}$, both positive \textit{and} negative. For a planar source emitting into one half of the space $z\!>\!0$, the components $k_{z}\!<\!0$ are incompatible with relativistic causation and propagation \cite{Yessenov19PRA}. An alternative is to produce the free-space O-shaped STWP via a moving point source comprising an optical dipole and sink (see \cite{Saari04PRE}, and more recently \cite{WilczekBook}). Consequently, all propagation-invariant wave packets reported to date in free space have been X-shaped \cite{Yessenov22AOP}. 

Anomalous GVD produces a curved light-cone surface [Fig.~\ref{Fig:Concept}(a)] such that the intersection of a plane tilted at the appropriate angle yields a closed elliptical spectral support that lies within the paraxial regime. This has enabled us recently to realize optical de~Broglie-Mackinnon wave packets that are subluminal ($\widetilde{v}\!<\!\widetilde{v}_{\mathrm{m}}$) \cite{Hall23NP}. The transition to X-shaped profiles only occurs in the superluminal regime after $\widetilde{v}\!>\!\widetilde{v}_{\mathrm{e}}$. By operating deep within the superluminal regime, we have been able to observe the X-to-O transition after tuning the group velocity through $\widetilde{v}_{\mathrm{e}}$. Note that the oppositely signed curvature of the light-cone in the normal-GVD regime precludes the possibility of sustaining O-shaped STWPs \cite{Malaguti09PRA,Hall23Normal}.

Finally, we note that observing the X-to-O-wave structural transition for a propagation-invariant STWP in a dispersive medium requires exquisite control over its spatiotemporal structure (in absence of added spectral-phase structure \cite{Kondakci18PRL,Yessenov19Optica,Wong21OE}). Such precision has not yet been realized in approaches that rely on nonlinear optics to produce conical waves \cite{DiTrapani03PRL,Faccio07OE}. Although a closed elliptical spatiotemporal spectrum can be enforced by the phase-matching conditions in presence of anomalous GVD, observation of the O-shaped profile of a propagation-invariant wave packet has not yet been reported, whether in the classical regime in nonlinear bulk media \cite{Porras05OL} or nonlinear multimode waveguides \cite{Stefanska23ACSP}, or in the quantum regime \cite{Spasibko16OL,Cupita20OL}. Our work here suggests that such nonlinear and quantum processes may be seeded by linearly synthesized STWPs endowed with prescribed spatiotemporal spectra, thus offering the potential for new opportunities in phase-matched nonlinear interactions. Examples include producing novel entangled-photon states via spontaneous parametric downconversion with designer-made spectral correlations between the two photons \cite{Spasibko16OL,Cupita20OL}, and producing high-energy STWPs in optical parametric amplifiers \cite{Li22CP}.

In conclusion, we have observed the long-predicted X-to-O structural field transition for propagation-invariant STWPs in a dispersive medium at a \textit{fixed} wavelength in presence of anomalous GVD. Previous attempts at observing this transition relied on changing the wavelength across the ZDW in the medium. In contrast, by tuning the group velocity of an STWP in the medium at a fixed wavelength, the spatiotemporal spectrum either re-intersects with the curved dispersive light-line (leading to a closed elliptical spectrum associated with an O-shaped profile), or escapes re-intersection (leading to an open hyperbolic spectrum associated with an X-shaped profile). These results and their underlying methodology open new avenues for exploring phase-matching in nonlinear media, with potential applications in quantum optics for synthesizing novel entangled-photon states, and in preparing high-energy STWPs.

\section*{Funding}
Office of Naval Research (N00014-17-1-2458, N00014-20-1-2789).
\section*{Data availability}
Data underlying the results presented in this paper are not publicly available at this time but may be obtained from the authors upon reasonable request.
\section*{Disclosures}
The authors declare no conflicts of interest.

\section*{Appendix}

\begin{figure}[t!]
\centering
\includegraphics[width=8.6cm]{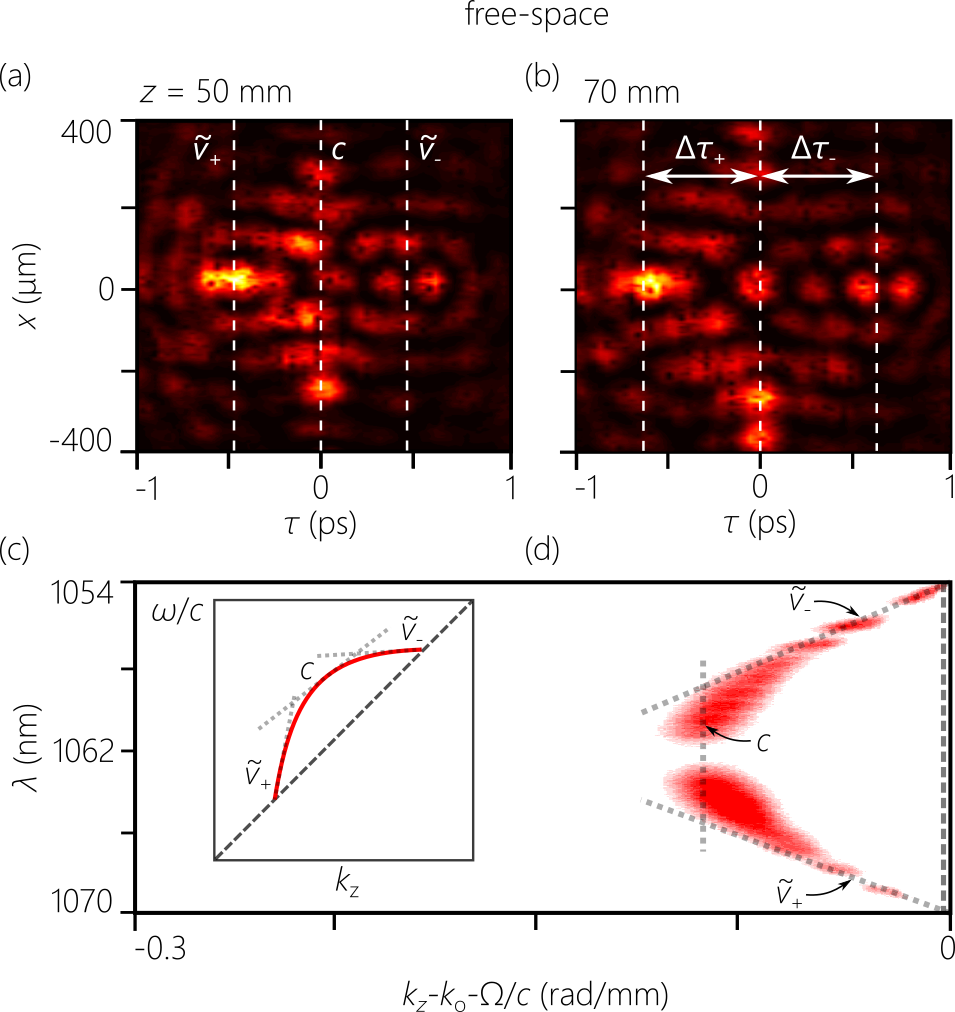}
\caption{Spatiotemporal profiles $I(x,z;\tau)$ for the O-shaped STWP in free space from Fig.~\ref{Fig:PropagationInvariance}(c) measured at (a) $z\!=\!50$~mm and (b) 70~mm. The vertical white lines identify the centers of three wave packets that undergo temporal separation (or walk-off) due to group-velocity mismatch. Initially at $z\!=\!0$, all three wave packets coincide. (c) The measured spatiotemporal spectral $(k_{z},\lambda)$-projection for the STWP in (a,b). The group velocities corresponding to the highlighted wave packets in (a,b) are identified. The inset shows the same spectral projection onto the $(k_{z},\tfrac{\omega}{c})$-plane schematically without modification of the horizontal axis.}
\label{Fig:PropagationInvariance2}
\end{figure}

We examine here in more detail the changes in the structure of the dispersive O-shaped STWP in free space [Fig.~\ref{Fig:PropagationInvariance}(c)], which is absent from its propagation-invariant counterpart in the dispersive medium. We plot in Fig.~\ref{Fig:PropagationInvariance2}(a,b) the spatiotemporal profile of this STWP at axial planes $z\!=\!50$~mm and 70~mm (the dispersion length in free space for this STWP is $\approx12$~mm). It is clear that the initial central peak in the spatiotemporal profile has split into three temporally separated peaks [Fig.~\ref{Fig:PropagationInvariance2}(a)], and that this separation increases with propagation distance [Fig.~\ref{Fig:PropagationInvariance2}(b)]. This indicates that the the O-shaped STWP in free space can be decomposed into three dispersive wave packets that travel at different group velocities, such that their walk-off produces the observed temporal separation in Fig.~\ref{Fig:PropagationInvariance2}(a,b). Because the profiles in Fig.~\ref{Fig:PropagationInvariance2}(a,b) are plotted in a frame moving at $c$ ($\tau\!=\!t-z/c$), we conclude that the stationary central peak corresponds to a luminal wave packet, whereas the leading and trailing wave packets are superluminal and subluminal, respectively, with group velocities estimated at $\widetilde{v}_{+}\!\approx\!1.004c$ and $\widetilde{v}_{-}\!\approx\!0.996c$.

This hypothesis is confirmed by examining the structure of the spatiotemporal spectrum projected onto the $(k_{z},\lambda)$-plane in free space [Fig.~\ref{Fig:PropagationInvariance2}(c)]. Whereas this spectral projection is a straight line in the dispersive medium (indicating dispersion-free propagation at a fixed group velocity), this projection in free space is curved [Fig.~\ref{Fig:PropagationInvariance2}(c)]. Three regions can be identified in this $(k_{z},\lambda)$-projection whose slopes correspond to the three group velocities mentioned above: a subluminal portion with $\widetilde{v}_{-}\!=\!0.996c$, a luminal portion with group velocity $\widetilde{c}$, and a superluminal portion with $\widetilde{v}_{+}\!=\!1.004c$. These three segments correspond to dispersive wave packets whose propagation dynamics is consistent with the field structure in Fig.~\ref{Fig:PropagationInvariance2}(a,b). The presence of anomalous-GVD `straightens' out this $(k_{z},\lambda)$-projection, leading to the entire spectrum contributing to a dispersion-free O-shaped STWP.

\bibliography{diffraction}

\end{document}